\documentclass[twoside, transmag]{IEEEtran}
\usepackage[english]{babel}
\usepackage{graphicx}
\usepackage{booktabs}
\usepackage{amsmath}
\usepackage[]{siunitx} 
\DeclareSIUnit\torr{Torr}	
\usepackage{array}
\usepackage{float}
\usepackage[compress]{cleveref}
\Crefname{figure}{Fig.}{Figs.} 
\usepackage[noadjust, nospace]{cite} 
\usepackage[]{placeins}	
\usepackage[caption=false, caption=false]{subfig}	
\usepackage{color}
\usepackage{soul}	

\usepackage{makecell} 

\newcommand{\executeiffilenewer}[3]{\ifnum\pdfstrcmp{\pdffilemoddate{#1}}{\pdffilemoddate{#2}}>0 %
{\immediate\write18{#3}}\fi %
}

\begin{document}

\newcommand{\titletext}{Reducing cavity-pulling shift in Ramsey-operated compact clocks}	

\title{\titletext}
\author{Michele Gozzelino, Salvatore Micalizio$^*$, Filippo Levi, Aldo Godone and Claudio E. Calosso}
\markboth{Journal of \LaTeX\,~Vol.~?, No.~?, March~2018}%
{Gozzelino \MakeLowercase{\textit{et al.}}: \titletext}

\maketitle
\begin{abstract}
We describe a method to stabilize the amplitude of the interrogating microwave field in compact atomic clocks working in a Ramsey approach. In this technique, we take advantage of the pulsed regime to use the atoms themselves as microwave amplitude discriminators.

Specifically, in addition to the dependence on the microwave detuning, the atomic signal after the Ramsey interrogation acquires a dependence on the microwave pulse area (amplitude times duration) that can be exploited to implement an active stabilization of the microwave field amplitude, in a similar way in which the Ramsey clock signal is used to lock the local oscillator frequency to the atomic reference. The stabilization allows to reduce the microwave field amplitude fluctuations which in turn impact the clock frequency through cavity-pulling. The proposed technique has shown to be effective to improve our clock frequency stability on medium and long term.
We demonstrate the method for a vapor-cell clock working with a hot sample of atoms but it can be  extended to cold-atom compact clocks.

\end{abstract}

\section{Introduction}
\label{sec:intro}
\IEEEPARstart{C}{
ompact} atomic clocks have considerably increased their performances in the last 20 years. Two approaches have been successfully investigated and developed. In the first one, the core of the clock apparatus is a sample of laser-cooled atoms. These clocks are relatively recent \cite{Esnault:2010, cyl_cav_coldRb:2015, CPT-cold-NIST:2013} and can in principle provide the accuracy information not only on ground applications but, taking advantage of the micro-gravity environment, also in space \cite{Salomon:2001}.

\
The second approach is based on a hot vapor cell where the atoms are pumped by a resonant laser \cite{Vanier_rev:2007, vapor-cells_review:2015}. The cell usually contains a buffer gas that induces a significant frequency shift and compromises the device accuracy. However, these clocks exhibit very interesting fractional frequency stability, in the short \cite{Bandi:2014, Yun:2017, Hao:2016, Hafiz:2017, Kang:2015} and in the long term \cite{Micalizio:2012}, while maintaining a compact and robust set-up.

\
In both approaches, many clock implementations adopt a pulsed regime of operation where the clock transition is probed through a Ramsey scheme \cite{Vanier:1989}. This technique produces narrow resonance lines with lower sensitivity to microwave amplitude fluctuations of the interrogating field. Nevertheless, the frequency of the clock transition (and thus the clock stability) is still sensitive in a residual form to microwave power fluctuations through several physical phenomena \cite{cav_pulling:2004, cav_pulling:2006}. These effects contribute to a shift of the clock frequency as a function of the microwave power seen by the atoms. One of these effects is cavity-pulling shift, which is due to the feedback of the cavity on the atoms, when the cavity itself is detuned from the atomic clock frequency \cite{Vanier:1989}. In order to reach a long term fractional frequency stability at the level of $10^{-15}$, an active amplitude stabilization of the microwave field interacting with the atoms is required.

We propose a method to stabilize the microwave field amplitude that uses the atomic sample as a discriminant. Techniques that involve an atomic sample as the field-amplitude probe have been demonstrated in continuous operation (making use of the so-called ``Rabi resonances'' \cite{Camparo:atomic-candle, Sun:2017}) and with time-domain measurements \cite{Horsley:2013} .
Here we present an approach suitable to the pulsed regime. In this case, the atomic population imbalance providing the clock signal directly depends on the microwave pulse area (amplitude times duration), through the Rabi oscillations. For this reason, the atomic signal naturally constitutes a very reliable probe of the magnetic field as seen by the atoms.
Usual ways of stabilizing the microwave field amplitude are based on the measurement of the power exiting the synthesizer. Neither information on the properties of the cavity and of the coupling system, nor the actual field-amplitude inside the cavity are retrieved in this manner. Conversely, the proposed method is mostly insensitive to the coupling system and allows to reach the optimal set-point for the clock operation. Notably, this method requires minimal hardware modifications to the usual set-up and does not lead to short-term stability degradation (see \Cref{sec:algorithm}), since it does not introduce dead-time for the main frequency loop.

The actual implementation and measurements presented refer to a pulsed optically pumped (POP) Rubidium clock, but the technique can be easily generalized to other systems based on the Ramsey interrogation, including cold atom frequency standards \cite{Cutler:2005}.

In \Cref{sec:setup} we shortly review the main characteristics of the POP clock. In \Cref{sec:uW_stab} we provide the description of the proposed microwave amplitude stabilization technique. In \Cref{sec:cav_pull_calc} we briefly introduce cavity pulling as the physical effect which makes the clock frequency sensitive to microwave field
amplitude fluctuations. We then describe an active stabilization technique to reduce cavity-pulling shift itself. Finally, in \Cref{sec:stability}, we present some experimental data that validate the effectiveness of the method.

\section{Clock setup and operation}
\label{sec:setup}
\begin{figure}[h]
	\begin{center}
		\includegraphics[width=\columnwidth]{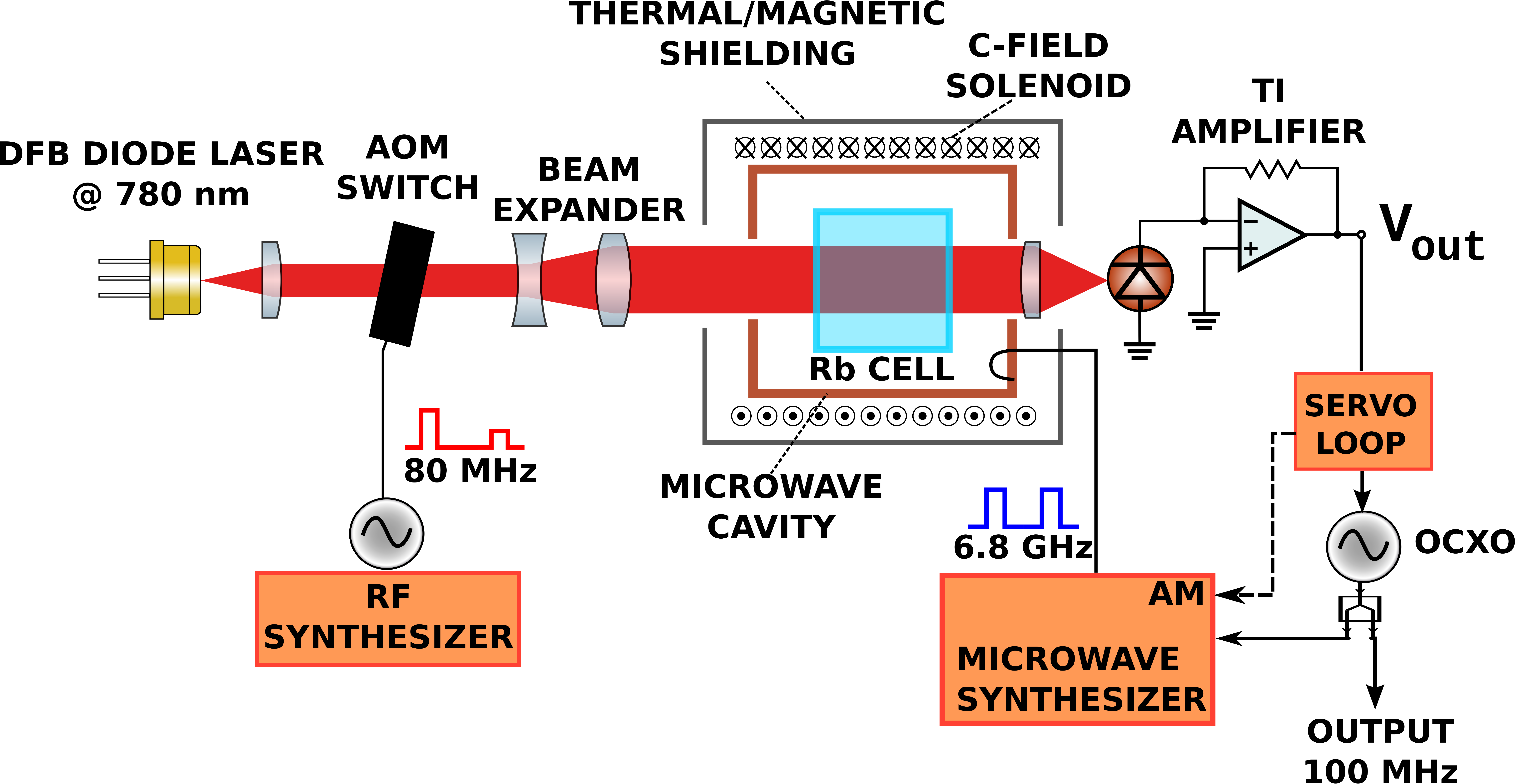}
	\end{center}
	\caption{Simplified scheme of the POP clock physics and optical packages. An FPGA provides the timing for the pulsed operation as well as the data acquisition and feedback loop.}
	\label{fig:setup}
\end{figure}

The clock setup is mainly composed of three subsystems: optics, physics package \cite{Micalizio:2012} and electronics \cite{Calosso:2007,Calosso:2017}. The basic scheme of the clock components is presented in \Cref{fig:setup}.

The optics includes a distributed feedback (DFB) diode laser tuned to the D$_2$ absorption line of $^{87}$Rb (780 nm); its frequency is locked to one of the sub-Doppler optical transition lines observed in an auxiliary reference cell. The main beam is linearly polarized and sent to the clock resonance cell through an acousto-optic modulator (AOM) and a beam expander.

The physics package stores the resonance cell filled with isotopically enriched $\mathrm{^{87}Rb}$ atoms and a temperature-compensated mixture of N$_{2}$-Ar buffer gas with total pressure
of approximately \SI[mode=text]{25}{\torr}. The cell is  placed inside a microwave cavity tuned to the Rb ground-state hyperfine splitting ($\omega_{12}$) and resonating on the TE$_{011}$ mode. The physics package is completed by magnetic and thermal shields, heaters and a solenoid providing the longitudinal quantization field (``c-field'').

The electronics drives all the clock phases: in particular it provides the RF pulses for the AOM-switching to pump and detect the atoms; it synthesizes the two low-noise phase-coherent microwave pulses at 6.834 GHz starting from a 100 MHz quartz oscillator; finally, it manages the digital acquisition and processing of the atomic signal \cite{Calosso:2007}.

\begin{figure}[h]
	\begin{center}
		\includegraphics[width=0.75\columnwidth]{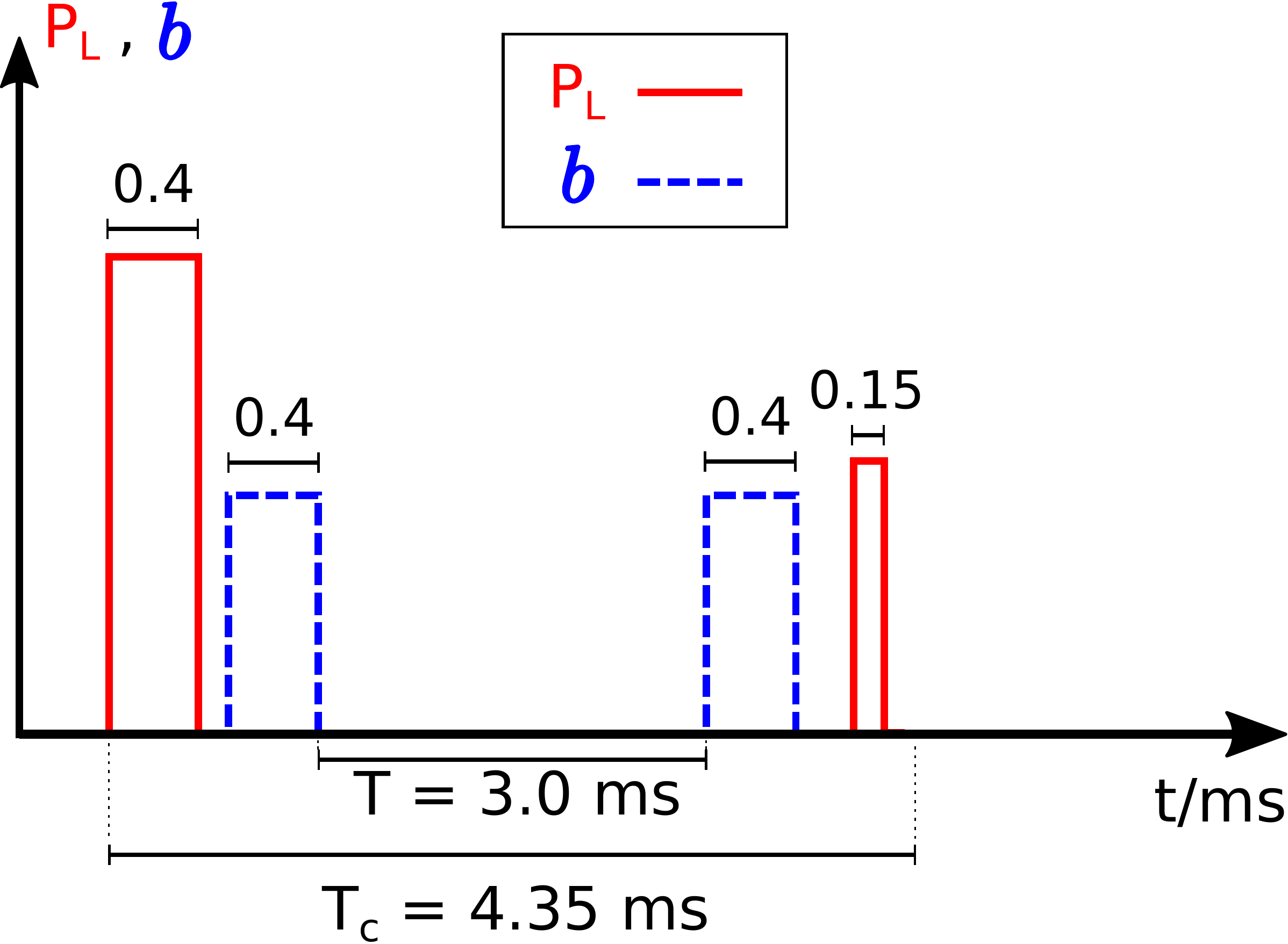}
	\end{center}
	\caption{Timing of the basic clock sequence which foresees temporally separated state preparation (by laser pumping), microwave clock interrogation and optical detection.}
	\label{fig:timing}
\end{figure}

The atomic signal is the result of a basic clock sequence that comprises atomic state preparation, clock interrogation and detection (see \Cref{fig:timing}). The preparation of Rb atoms in one of the two hyperfine levels is obtained by laser pumping as the AOM is driven at maximum power; the interrogation is carried out with the Ramsey technique through two temporally-separated microwave pulses with frequency $\omega_0$ and area $\theta= b t_{1}$ (where $b$ is the effective amplitude\footnote{$b$ is the actual field amplitude only if the field is constant over the whole area of the resonance cell. In case of non-uniformity of the field, it has the meaning of effective amplitude $b_e = \int_i{f(x,y,z)b_0} dV$, where $i$ is the interaction volume, $b_0$ is the amplitude at the center of the cavity and $f$ is a weighting function. The proposed method has the advantage of discriminating $b_e$ by maximizing the atomic signal, independently of the function $f$.} of the microwave field expressed in terms of Rabi frequency and $t_{1}$ is the pulse duration).
Finally, the clock transition is probed by driving the AOM with a weak RF pulse and detecting the transmitted light through a silicon photodiode and a transimpedance amplifier.

\section{Microwave field amplitude stabilization: method}
\label{sec:uW_stab}
\subsection{Error signal construction}
 The signal $S$, obtained integrating the transimpedance output over the detection window, for small frequency detuning ($\omega_0 - \omega_{12} \ll 1/t_1$) and for $\theta \simeq \pi/2$, can be modelled by the following (simplified) relation \cite{Vanier:1989}:
\begin{equation}
\label{eq:Signal}
S(\omega_0, \theta) \propto \sin^2\theta \left[1- \cos\left(\omega_0-\omega_{12}\right)T\right]
\end{equation}
where $T$ is the Ramsey time and the DC background is dropped. From now on, both $b$ and $\theta$ are intended to be representative of the microwave field amplitude, since $t_1$ is kept constant during the experiments and, considering the time stability of the FPGA that generates it~\cite{Calosso:2014}, it is at least two orders of magnitude more stable than $b$ (the FPGA clock is directly derived from the OCXO).

\begin{figure}[h]
	\centering
	\includegraphics[width=\columnwidth]{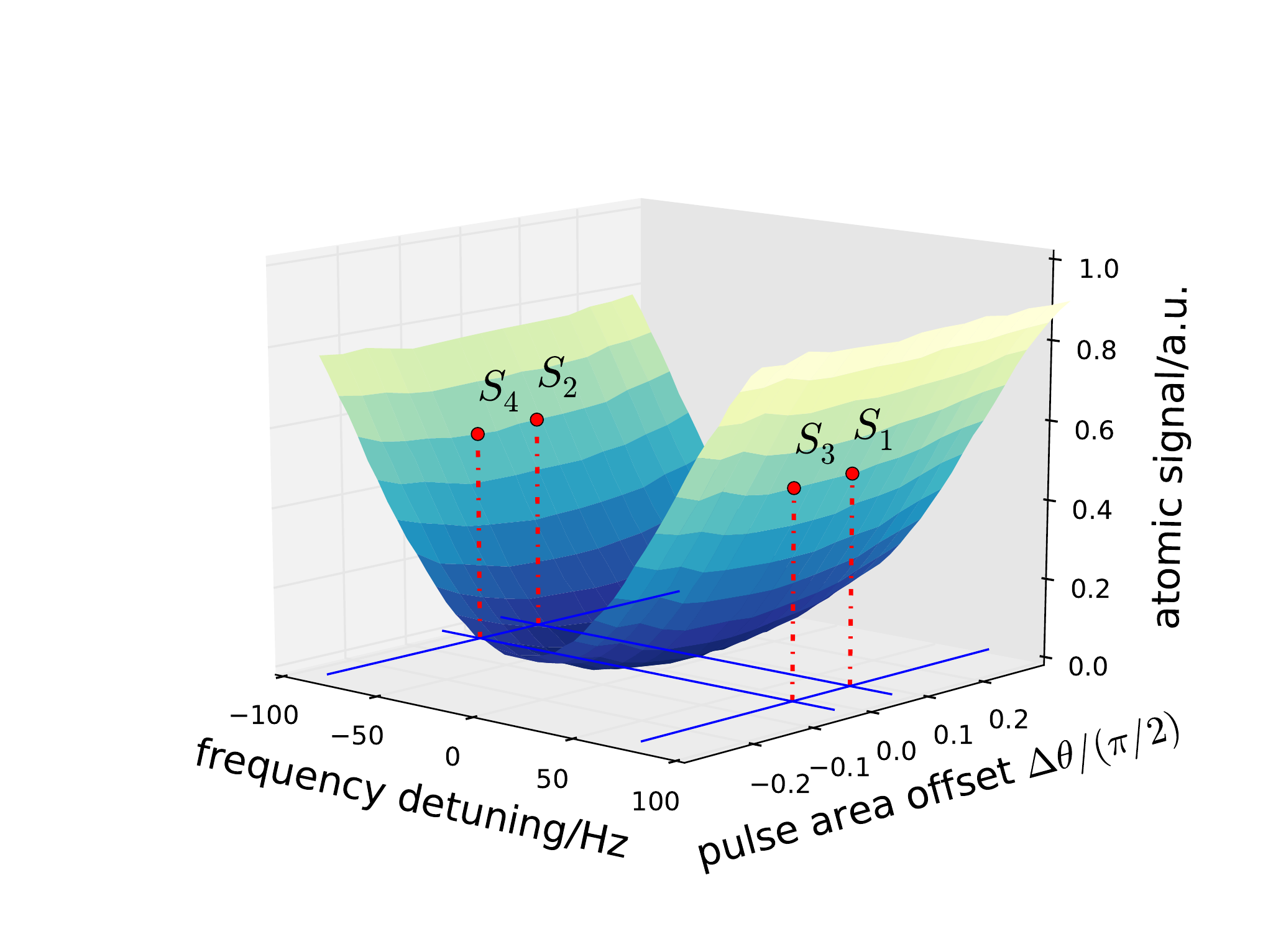}
	\caption{Atomic signal as a function of both the frequency detuning $(\omega_0-\omega_{12})/2\pi$ and the pulse area (or, equivalently, the pulse amplitude ($b$) at fixed pulse length $t_1$). The experimental data used to draw the plot are the result of several frequency scans at different fixed power levels (3 average for each scan). The  points labelled $S_i$ indicate the sampling points used in the extended method (see \Cref{sec:algorithm}). }
	\label{fig:signal3D}
\end{figure}

\Cref{fig:signal3D} shows an experimental scan of the signal $S$ in the two variables. It is clearly visible that a local minimum is present in both the dimensions, for $\omega_0 = \omega_{12}$ and $\theta = \pi/2$.
As a function of $\omega_0$, equation \labelcref{eq:Signal} reproduces the well known sinusoidal behavior of the central  Ramsey fringe. In normal operation, $\theta$ is set around $\pi/2$ in order to maximize the visibility of the fringes. A dispersive error signal is obtained by differentiating the output with respect to microwave frequency: the atoms are investigated at $\omega_1 = \omega_0 + \omega_m$ and at $\omega_2 = \omega_0 - \omega_m$; the related signals ($S_1$ and $S_2$ respectively) are then used to construct the error signal $E \equiv S_2-S_1$. The latter is fed to the frequency controller to calculate the correction $C$ to be applied to the local oscillator. When the frequency loop is closed, the error signal is nulled and the microwave frequency is locked to the minimum of the central Ramsey fringe ($\omega_0 = \omega_{12}$). $\omega_m$ is chosen to be at the fringe half-width half-maximum (HWHM) in order to maximize the slope of the frequency discriminator.
As a function of $\theta$, in the $\sin^2\theta$ term of \cref{eq:Signal} we can recognize the well known Rabi oscillations \cite{Vanier:1989} that lead to a minimum of the transmission signal in correspondence to the working point ($\theta = \pi/2$). We can think to apply an amplitude modulation to detect and to compensate for amplitude variations of the microwave field, in analogy of what is usually done for the microwave frequency. This basic insight is the key-point of the proposed method, as addressed in the following section.

\subsection{Stabilization algorithm}
\label{sec:algorithm}
The microwave field amplitude stabilization is implemented as an extension of the usual clock operation. The microwave frequency and amplitude are simultaneously dithered and the atomic signal is used to close both the frequency and the amplitude stabilization loops. In order to differentiate the signal with respect to both microwave frequency and amplitude, four points are necessary, as described in \Cref{tab:sequence}. There, we refer to the microwave amplitude modulation depth as $b_m$.
\begin{table}[h]
					
\caption{Sequence for the $n-$th clock cycle which includes the frequency and amplitude modulation}
		\centering 
		\resizebox{0.45\textwidth}{!}{
		\begin{tabular}[]{m{4 em} c c c} 
				\toprule   
					&  \thead{microwave \\ frequency} & \thead{microwave pulse\\ amplitude} & \thead{Atomic\\ signal}		\\ 				
				\toprule[0.12em] 
			step 1 & $\omega_{0,n}-\omega_m$	 	&	$b_{0,n}-b_m$	&	$S_{1,n}$\\	
			step 2 & $\omega_{0,n}+\omega_m$	 	&	$b_{0,n}-b_m$ &	$S_{2,n}$\\	
			step 3 & $\omega_{0,n}-\omega_m$	 	&	$b_{0,n}+b_m$ &	$S_{3,n}$\\	
			step 4 & $\omega_{0,n}+\omega_m$	 	&	$b_{0,n}+b_m$ &	$S_{4,n}$\\	

			\bottomrule 
			\end{tabular}}
				\label{tab:sequence} 
\end{table}
The four steps of \Cref{tab:sequence} compose an ``extended'' clock cycle that is repeated indefinitely. For the $n-$th clock cycle, two error signals $E^{\omega}_{n}$ and $E^b_n$, proportional to $\omega_0 - \omega_{12}$ and to $\theta - \pi/2$ respectively, are generated:
\begin{align}
E^{\omega}_{n} & = (S_{4,n}+S_{2,n})-(S_{3,n}+S_{1,n}) \label{eq:E_omeg}\\
E^{b}_{n} & = (S_{4,n}+S_{3,n}) -(S_{2,n}+S_{1,n}) \label{eq:E_b}
\end{align}
The corrections for the local oscillator $C^\omega_n$ and for the microwave amplitude $C^b_n$ are provided by two pure integrative controllers:
\begin{align*}
C^{\omega}_n & = -k^{\omega} \sum_{j=1}^{n} E^{\omega}_j = C^{\omega}_{n-1} -  k^{\omega}E^{\omega}_n \\
C^{b}_n & = -k^{b} \sum_{j=1}^{n} E^{b}_j = C^{b}_{n-1} - k^{b}E^{b}_n
\label{eq:corrections}
\end{align*}

where $k^{\omega}$ and $k^{b}$ are multiplicative constants which are proportional to the gain and the bandwidth of the frequency and amplitude loops respectively.

The parameters of the controllers are set in order to minimally perturb the classical operation of the clock, which has been optimized for best performance. As usual, $\omega_m$ is kept at HWHM, whereas $b_m$ is a trade-off for having a sufficient sensitivity to amplitude variations without significant degradation of the frequency sensitivity. The low sensitivity to amplitude variations is compensated by a long time-constant of the amplitude loop that allows to integrate the signal for a long time (about 10 s). This is possible since amplitude fluctuations are slow, and impact the clock stability for averaging time longer than 1000 s. On the opposite, $k^\omega$ is as high as possible, with a time constant of the order of 100 ms.

This method does not have any impact to the Dick-effect contribution to the short-term stability \cite{Santarelli:1998, Presti:1998}. Even thought two additional basic sequences are introduced, the frequency-loop duty cycle remains unchanged, and the frequency information is acquired and stored in each clock sequence. In this sense, such a scheme is advantageous with respect to an interleaved and separate measurement of the frequency and amplitude variations, because dedicated amplitude investigations would behave as dead-time for the main frequency loop.

Amplitude corrections are experimentally implemented by acting on the AM input of the microwave synthesizer, requiring minimal hardware modifications to the basic setup (see dashed line in \Cref{fig:setup}). In detail, a 16-bit digital-to-analog converter (DAC) tunes by up to $\pm$15\% the full-scale current of the DDS of the synthesis chain, leading to an AM resolution of about 5 ppm \cite{Calosso:2007}.

\section{Cavity pulling reduction}
\label{sec:cav_pull_calc}
The major contribution to the sensitivity of the atomic clock frequency to microwave amplitude fluctuations comes from the cavity-pulling effect. When the cavity is detuned from the atomic frequency, the cavity feedback on the atoms induces a dephasing of the hyperfine coherence evolution during the Ramsey time. The knowledge and the control of this shift is of great importance both for cold-atom clocks and vapor-cell compact clocks. For the first ones, it has to be characterized in order to assess the accuracy budget. For the latter, the absolute value of the shift is not a concern, since accuracy is not needed. Nevertheless, it is of great importance to evaluate the system sensitivity to environmental parameters which can cause a variation in the shift, having an impact on the ultimate clock stability. In both cases, it is often needed to stabilize some parameters of interest to obtain the required frequency accuracy and stability. In this section we analyse how the microwave amplitude active stabilization can mitigate cavity-pulling effects. This can help to either reach better performances or to relax the requirements in the design of the physics package.

Several parameters affect the cavity-pulling shift, including the loaded quality factor of the microwave cavity ($Q_L$), the cavity detuning from the atomic resonance ($\Delta\omega_C = \omega_C-\omega_{12}$), the
microwave pulse area ($\theta$) and the Ramsey time $T$. In the limit of small cavity detuning ($\Delta \omega_{C} \ll \omega_{12}/2Q_{L}$), it is possible to find an analytical expression for the cavity-pulling shift ($\Delta \omega_{cp}$) of the central Ramsey fringe (see the appendix for the full expression):

\begin{equation}
\Delta \omega_{cp}= - \frac{2 Q_{L}}{T\omega_{12}}f(\theta, Q_{L}, T) \Delta\omega_{C}
\label{eq:cav_pull}
\end{equation}

where $f(\theta, Q_{L}, T)$ is a function of the microwave pulse area which crosses zero for $\theta\approx\pi/2$. We point out that $f(\theta, Q_{L}, T)$ also includes a not trivial dependence on the cavity quality factor $Q_L$ \cite{cav_pulling:2006}.

$Q_L$ and $\Delta\omega_C$ can change during the years of operation due to cavity aging. The main processes involved are mechanical relaxations of the cavity walls and the deposition of metallic layers upon impurities of the vapor cell walls \cite{Patton:2016}. The first process can mainly lead to a frequency detuning of the cavity, due to a change in the geometrical configuration. The metallic deposition in a region where the electric and magnetic fields are not negligible can also lead to a frequency detuning of the cavity resonance \cite{Pozar}. Moreover, the metallic layer will lead to augmented losses, by changing the total impedance of the cavity, and eventually change the total $Q_L$ \cite{Godone:2011}. With this in mind, it is reasonable to assume $Q_L$ decreasing over time as the metallic losses increase, whereas the behavior of $\Delta\omega_C$ cannot be predicted a priori but a (positive or negative) linear trend is supposed.

\begin{figure}[h]
	\centering
	\includegraphics[width=0.45\textwidth]{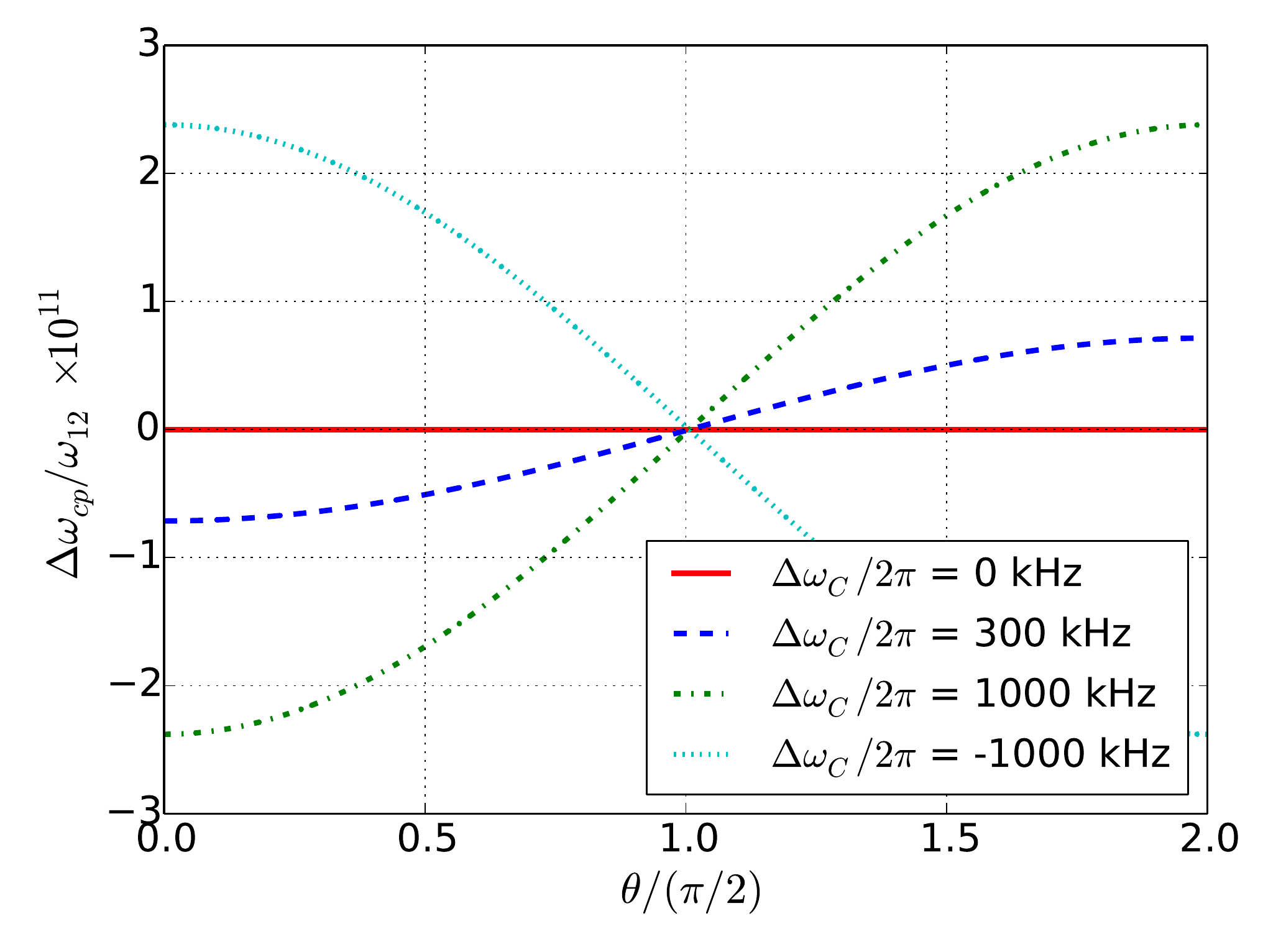}
	\caption{Calculated cavity pulling shift~\cite{cav_pulling:2006} as a function of $\theta$, for different values of cavity detuning and fixed Q-factor ($Q_L = 500$).}
	\label{fig:cav_pull_d}
\end{figure}

\begin{figure}[h]
	\centering
	\includegraphics[width=0.45\textwidth]{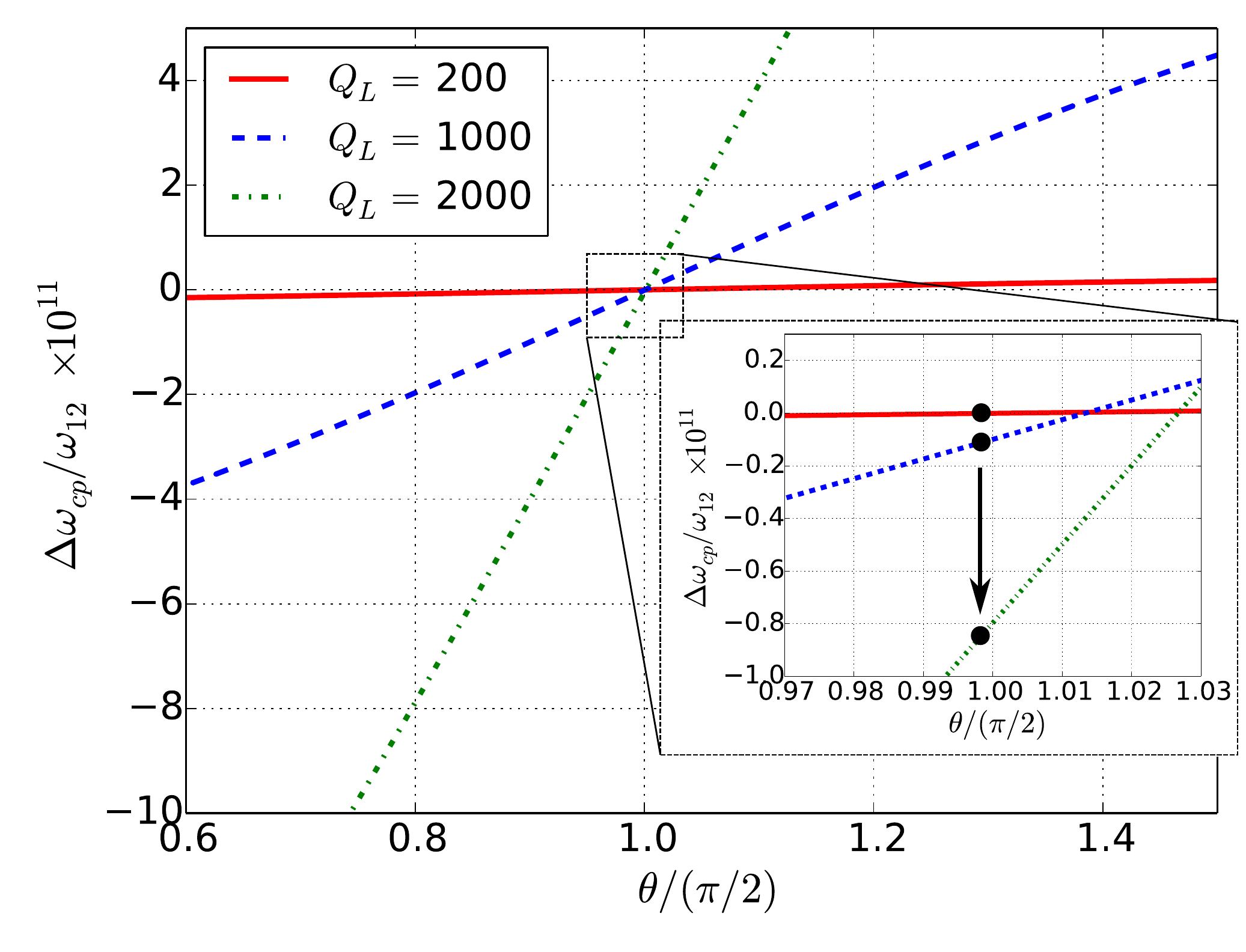}
	\caption{Calculated cavity pulling shift~\cite{cav_pulling:2006} as a function of $\theta$, for different values of the cavity loaded Q-factor and fixed detuning ($\Delta\omega_C = 2\pi\times 500$ kHz).}
	\label{fig:cav_pull_Q}
\end{figure}

In the following figures, we plot the calculated cavity pulling shift $\Delta\omega_{cp}$ from \labelcref{eq:cav_pull} versus the microwave pulse area $\theta$. In \Cref{fig:cav_pull_d} the shift is calculated for different values of $\Delta\omega_{C}$ at a fixed $Q_{L}$ and, similarly, in \Cref{fig:cav_pull_Q} for different $Q_{L}$ at a fixed $\Delta\omega_{C}$.
The slopes of the curves around the working point $\theta=\pi/2$ set the sensitivity of the clock frequency to microwave field amplitude fluctuations.
Specifically, figures \ref{fig:cav_pull_d} and \ref{fig:cav_pull_Q} show that the two main cavity parameters affect the clock frequency in different manner. In the first case, an augmented detuning $\Delta\omega_{C}$ raises the magnitude of the sensitivity, leaving the zero-crossing of the frequency shift unchanged.
If instead the loaded $Q$-factor gets lower with time, the sensitivity decreases but at the same time the frequency-shift zero crossing point changes (see \Cref{fig:cav_pull_Q} and inset therein). Thus a slow change of the $Q$-factor may induce a slow change of the clock frequency, even if the microwave amplitude is kept fixed by an active stabilization.
Before looking at the impact on the clock stability when applying the microwave stabilization, to give more insight into the capability of the active stabilization to improve the clock's performances, we discuss a numerical example, referring to literature for typical temporal variations of the cavity quality factor and resonant frequency.

Concerning the behavior of $Q_L$, \cite{Coffer:2004} reports a decay of the $Q$ factor of the form:
\begin{equation}
Q_{L}(t)=Q_{0}+Q_{1} e^{-t/\tau}
\label{eq:temporal_Q}
\end{equation}
The measured value of $Q$ is initially 2800 decreasing with a time constant of 67 days, thus reaching a stationary value of 1500. In our work \cite{Godone:2011}, a similar behavior is observed: however, the initial cavity $Q_L$ was of the order of 10000 decreasing with a much longer time constant, of the order of 3300 days. In the latter work, however, the clock transition was detected through the maser emission and the condition a high cavity $Q$-factor was mandatory to observe a significant signal-to-noise ratio. In the optical detection approach, a high $Q_L$ is not needed and we assume an initial value of 2000, slightly larger than that we measured more recently in our current set-up ($\approx 1000$). Without losing in generality, we adopt a time constant intermediate between the two values found in the literature and we assume $\tau= 200$ days. We also choose $Q_{0}=450$ (asymptotic value) and $Q_{1}=1550$ in the following calculation.

Concerning the mode resonance frequency, no change was observed in \cite{Coffer:2004}, while we measured a drift of \SI[mode=text]{-1}{\kilo\hertz}/day over a period of 3 years \cite{long_term:2010}. We keep this value in our simulation.

To estimate the magnitude of these effects on the clock frequency, we initially consider the situation where  the microwave amplitude stabilization scheme is not active. The fractional change of the cavity pulling shift due to (small) fractional variations of the microwave amplitude and cavity parameters is then given by:

\begin{equation}
	\frac{\delta(\Delta\omega_{cp})}{\omega_{12}} = \alpha \frac{\delta\theta}{\theta}+ \beta \frac{\delta Q_L}{Q_L}+ \gamma \frac{\delta \Delta\omega_{C}}{\Delta\omega_{C}}
	\label{eq:drift_OL}
\end{equation}

where the coefficients are defined as:

\begin{subequations}
	\label{eq:coeffs}
	\begin{align}
\alpha & =\frac{\theta}{\omega_{12}}\frac{\partial\Delta\omega_{cp}}{\partial\theta}\Bigg|_{Q_L, \Delta\omega_C}\\
\beta & =\frac{Q_{L}}{\omega_{12}}\frac{\partial\Delta\omega_{cp}}{\partial Q_{L}}\Bigg|_{\theta, \Delta\omega_C}\\
\gamma & =\frac{\Delta\omega_{C}}{\omega_{12}}\frac{\partial\Delta\omega_{cp}}{\partial \Delta\omega_{C}}\Bigg|_{\theta, Q_{L}}
	\end{align}
\end{subequations}

Due to the dependence of $Q_L$ and $\Delta\omega_C$ on time (see \labelcref{eq:temporal_Q} and text below), we do not expect the coefficients in equation (\labelcref{eq:coeffs}) to be constant; rather, they will depend on the time they are evaluated.
For example, after six months of operation we take $Q_{L}\approx 1000$, $\Delta\omega_{C} = $ \SI[mode=text]{-650}{\kilo \hertz}, thus (\labelcref{eq:coeffs}) and \labelcref{eq:cav_pull} yield:

\begin{align}
\label{eq:coeff_vals}
\alpha & \simeq -1.1 \times 10^{-10} \nonumber \\
\beta & \simeq	4.8 \times 10^{-12} \\
\gamma & \simeq 	1.6	\times 10^{-12}	\nonumber
\end{align}

where the coefficients are evaluated for $\theta=\pi/2$.
Taking as a reference the relative amplitude fluctuations measured in our frequency synthesis chain $\delta \theta/(\theta \Delta t)\approx 10^{-4}$/day \cite{long_term:2010}, from \labelcref{eq:drift_OL} we get an induced frequency-aging rate of the order of $10^{-14}$/day. If the variation in $Q_L$ and $\Delta\nu_C$ are reasonably under control (below \SI[mode=text]{1}{\percent}/week) the cavity pulling shift is dominated by microwave amplitude fluctuations, being $\alpha\delta\theta/\theta$ the dominant term.

\begin{figure}[h]
\centering
\subfloat[]{%
	\label{fig:temporal_det}	
  	\includegraphics[clip,width=0.8\columnwidth]{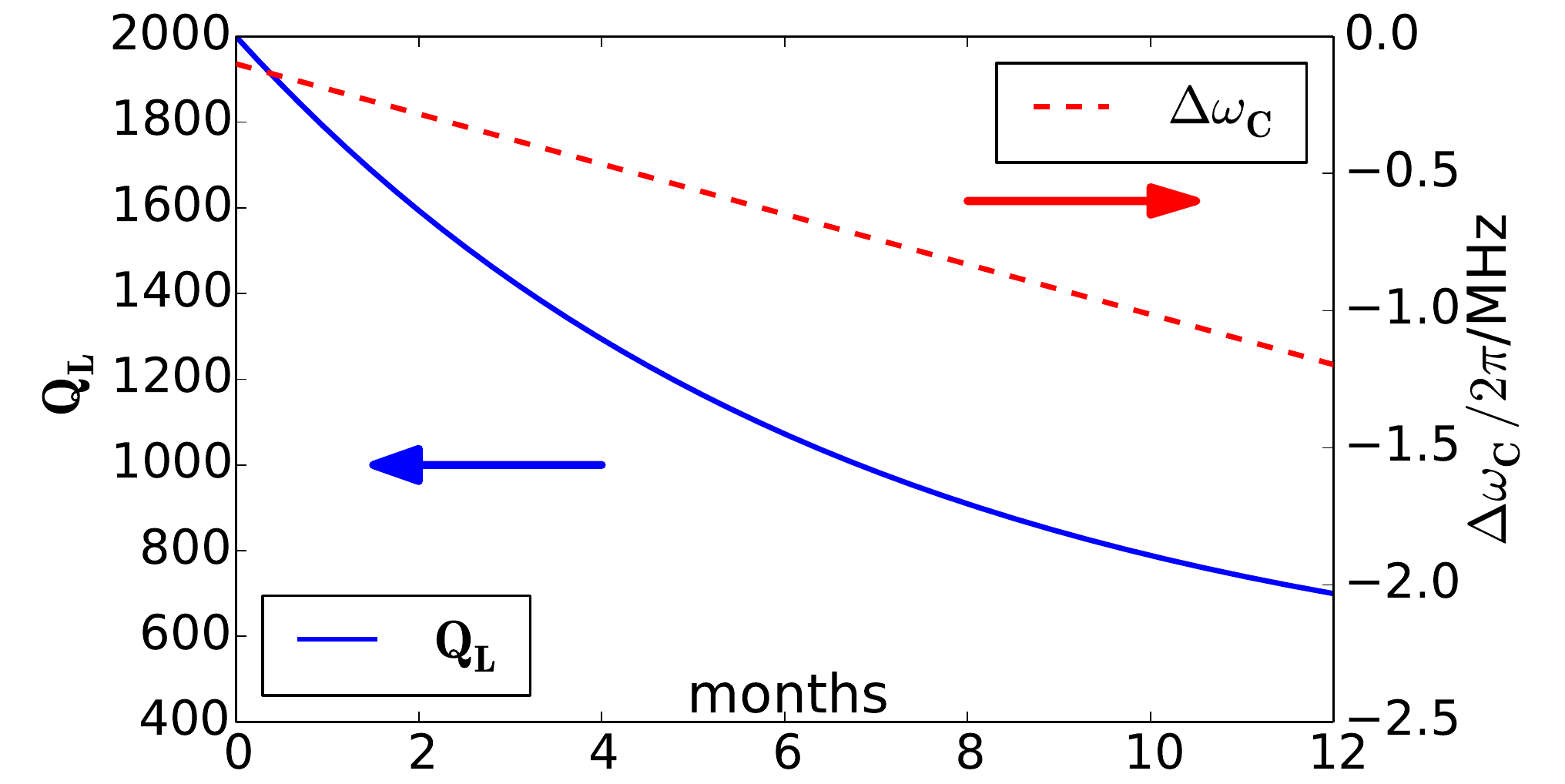}%
}
\hspace{-0.1cm}
\subfloat[]{%
	\label{fig:temporal_Q}	
	\includegraphics[clip,width=0.9\columnwidth]{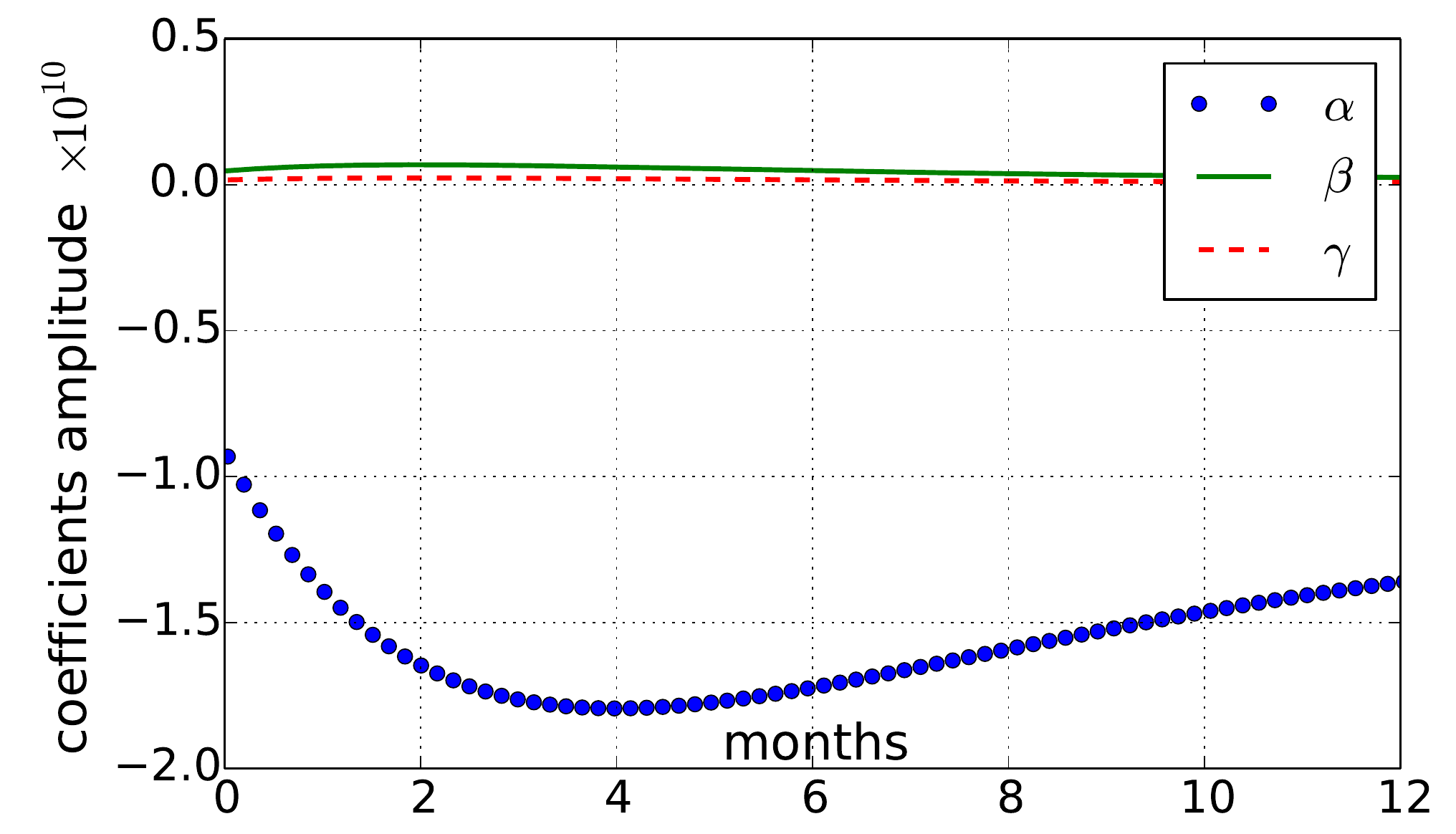}%
}
\hspace{-0.1cm}
\subfloat[]{%
	\label{fig:daily_df}	
	\includegraphics[clip,width=0.9\columnwidth]{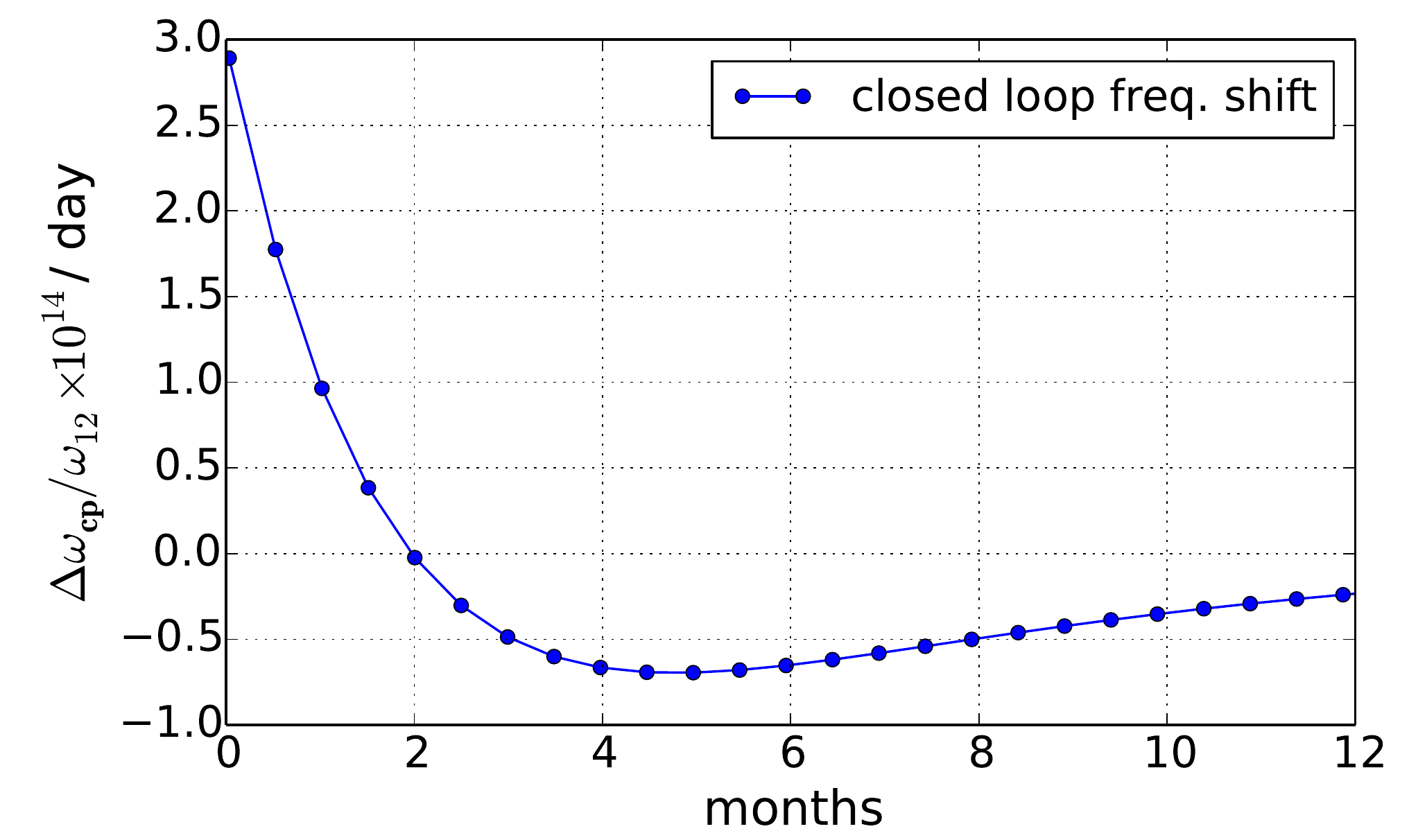}%
}

\caption[]{ \subref{fig:temporal_det} Detuning and $Q$-factor evolutions during a one-year period as assumed in the model calculations.  \subref{fig:temporal_Q} Sensibility coefficients of eq. (\labelcref{eq:coeffs}) during the same period.  \subref{fig:daily_df} Daily frequency variation, with pulse area actively stabilized at $\pi/2$, due to a change in the cavity pulling shift (eq. \labelcref{eq:drift_CL}) as $Q_L$ and $\Delta\omega_C$ change during one year.}

\label{fig:daily}
\end{figure}

However, by adopting the amplitude microwave stabilization technique, $\delta\theta/\theta$ is made negligibly small as amplitude fluctuations are corrected by the feedback loop and the frequency-aging rate turns out exclusively limited by the variation of the cavity parameters:
\begin{equation}
	\frac{\delta(\Delta\omega_{cp})}{\omega_{12}} = \beta \frac{\delta Q}{Q}+ \gamma \frac{\delta \Delta\omega_{C}}{\Delta\omega_{C}}
	\label{eq:drift_CL}
\end{equation}
As mentioned before, this frequency-aging rate will change with time, if the variation in the cavity parameters is not stationary. This effect is shown for the model calculation of \Cref{fig:daily}.
It is evident from the figure that, aside the initial transient where $Q_L$ is changing quite fast, the cavity-pulling induced frequency-aging can be constrained below $10^{-14}$/day when the active amplitude stabilization method is implemented (blue dotted line). An optimal configuration to further reduce the sensitivity to the last two coefficient would be to add an auto-tuning system \cite{Audoin:1981, Godone:2011}, of course at the expense of adding microwave components thus increasing the complexity of the set-up.

\section{Long-term stability improvement}
\label{sec:stability}
In this section we validate the performances of the double loop and we compare the stability performances of the previously described POP clock with and without active control of the microwave field amplitude. The measurements presented in this section were obtained with a frequency modulation depth of $\omega_m = 2\pi\times80$ Hz and a microwave amplitude modulation depth of $b_m = 0.05 \; b_0$. The timing is the one reported in \cref{fig:timing}. The rest of the settings are the same as in \cite{Micalizio:2012}. The cavity quality factor is around 500 and the cavity detuning is of the order of -1 MHz. The POP clock was measured against a reference oscillator, which is a composed of a BVA option 08 phase-locked to an active hydrogen maser (Kvarz ch1-75). Between the two measurements, we reduced the time constant of the phase-lock loop (from \SI[mode=text]{3}{\second} to \SI[mode=text]{100}{\milli\second}), to improve the frequency stability of the reference. All other experimental parameters are the same in the two configurations. As we will see later, this change has an impact only in the short-term stability evaluation, thus it is not of concern for the comparison.

\begin{figure}[h]
\centering
\subfloat[]{%
	\label{fig:valid_a}	
  	\includegraphics[clip,width=0.95\columnwidth]{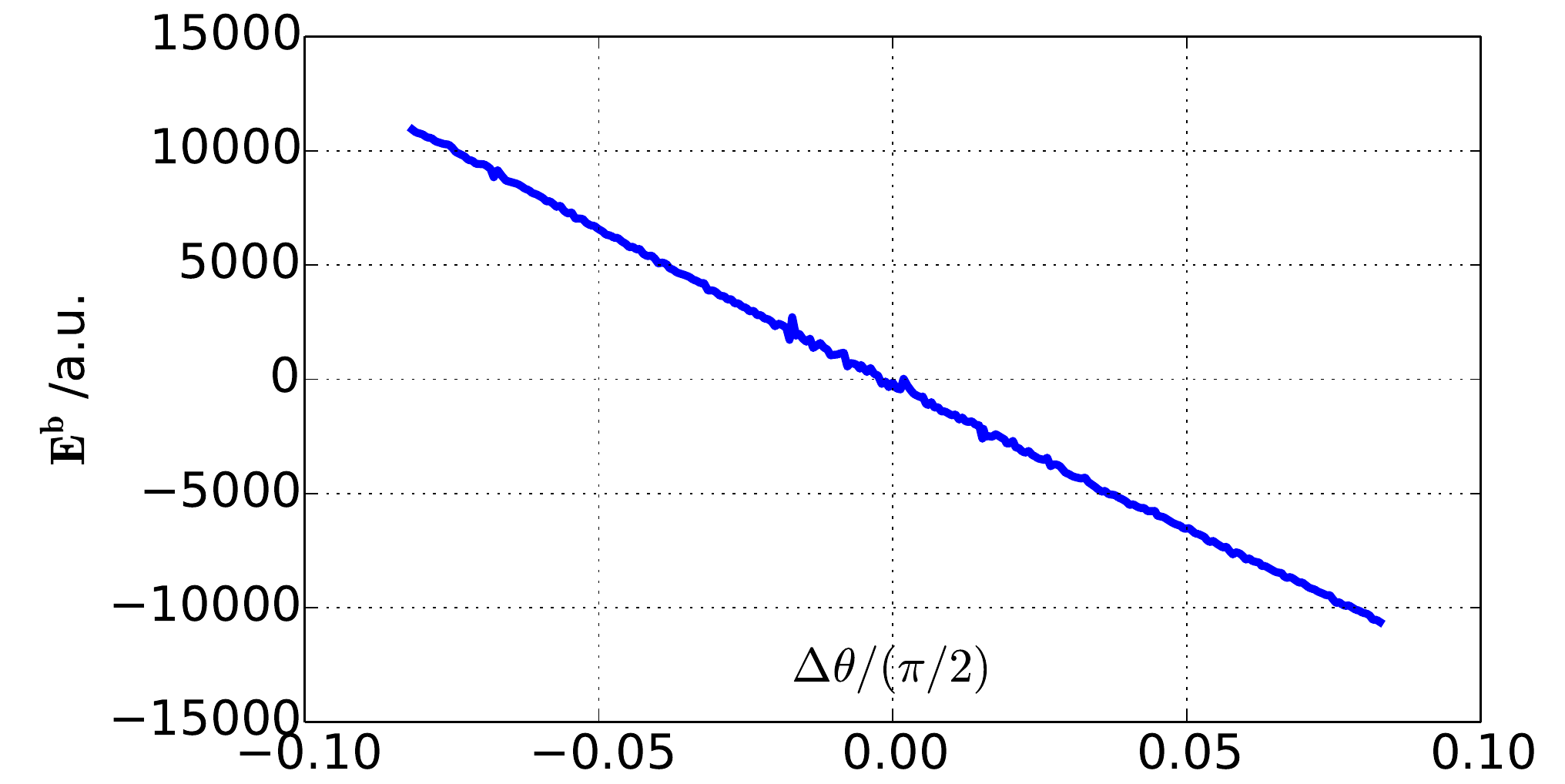}%
}
\hspace{-0.1cm}
\subfloat[]{%
	\label{fig:valid_b}	
	\includegraphics[clip,width=0.95\columnwidth]{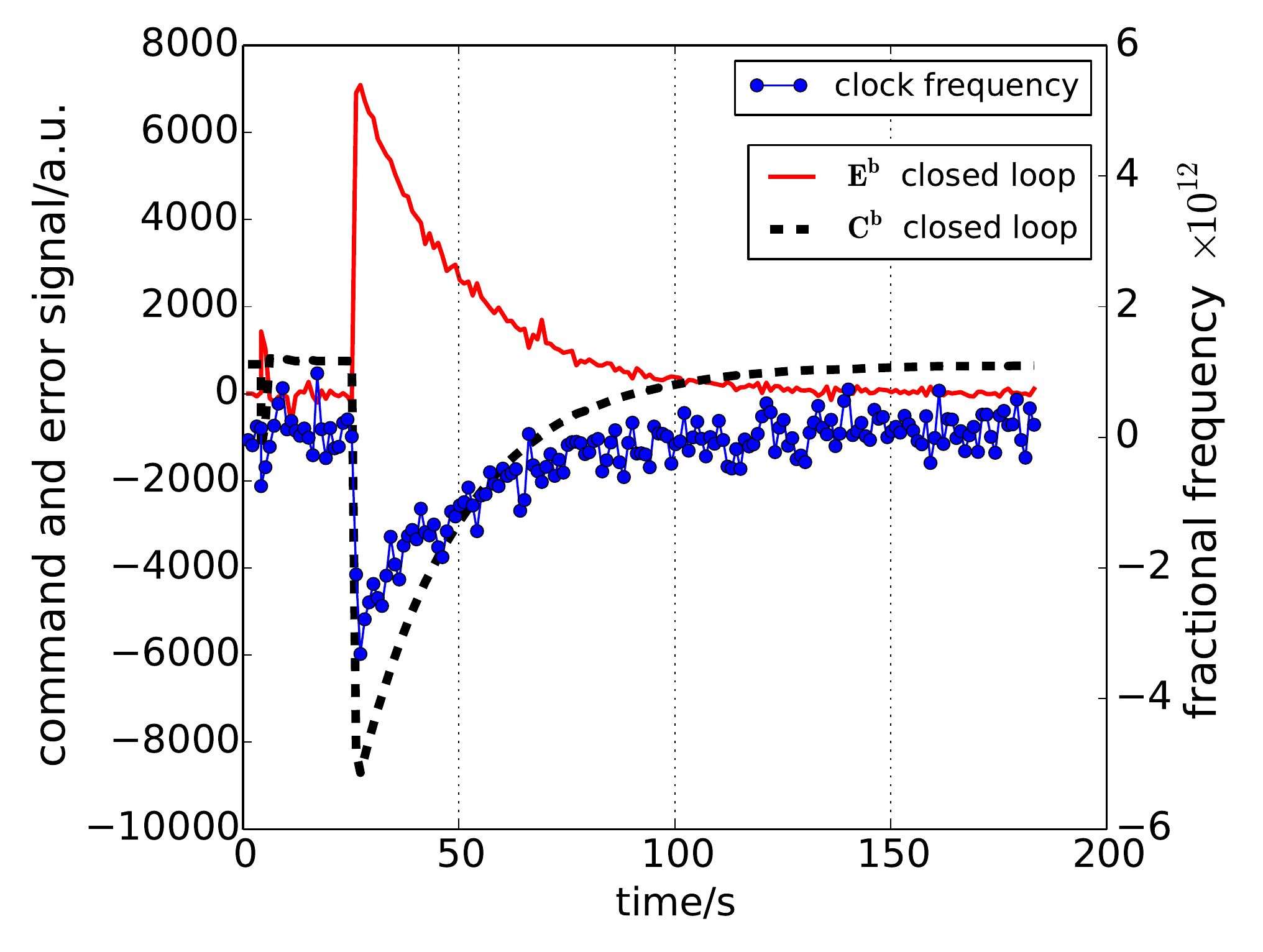}%
}
\hspace{-0.1cm}

\caption[]{ \subref{fig:valid_a} Microwave amplitude error signal $E^b$ in open loop as a function of the pulse area offset $\Delta\theta$. \subref{fig:valid_b} In-loop error signal $E^b$ (continuous line) and correction signal $C^b$ (dashed line) as an arbitrary step in the pulse area offset is introduced (left \textit{y}-axis). Fractional frequency variation synchronously acquired (right \textit{y}-axis, dotted line).}

\label{fig:validate}
\end{figure}

With the frequency loop closed, we validate the amplitude stabilization method by measuring the error signal $E^b$ versus the pulse-area offset $\Delta\theta$ (Fig. \ref{fig:valid_a}) in open amplitude loop. We observed a sensitivity of the discriminator of 1300/\%. We believe that the magnitude of the discriminant slope is given by two factors: the sharpness of the Rabi oscillations peak around $\pi/2$ and the amplitude modulation depth $\theta_m$. Well defined Rabi oscillations have been observed for a loaded $Q_L$ as low as 100, therefore the proposed method is suitable to common cavity-cell arrangements \cite{Bandi:2014, Horsley:2013}. Moreover, we acquired $E^b$, $C^b$ and the clock fractional frequency with both loop closed, in response to an externally imposed negative step of the microwave field amplitude (Fig. \ref{fig:valid_b}). The negative step in closed loop induces a frequency jump of $4\times10^{-12}$ in the clock frequency that is recovered by the loop with a time constant $\tau_L$ of 25 s.
In general, the microwave fluctuations expressed as AM power spectral density are reduced by $(f/B_L)^2$, where $B_L=\frac{1}{2\pi\tau_L}$ is the bandwidth of the loop. In terms of Allan deviation, the improvement is $k\frac{\tau_L}{\tau}$, where $k$ is a multiplicative constant that depends on the noise type and equals $\sqrt{3}$ in case of random walk.

\begin{figure}[h]
	\centering
	\includegraphics[width=0.5\textwidth]{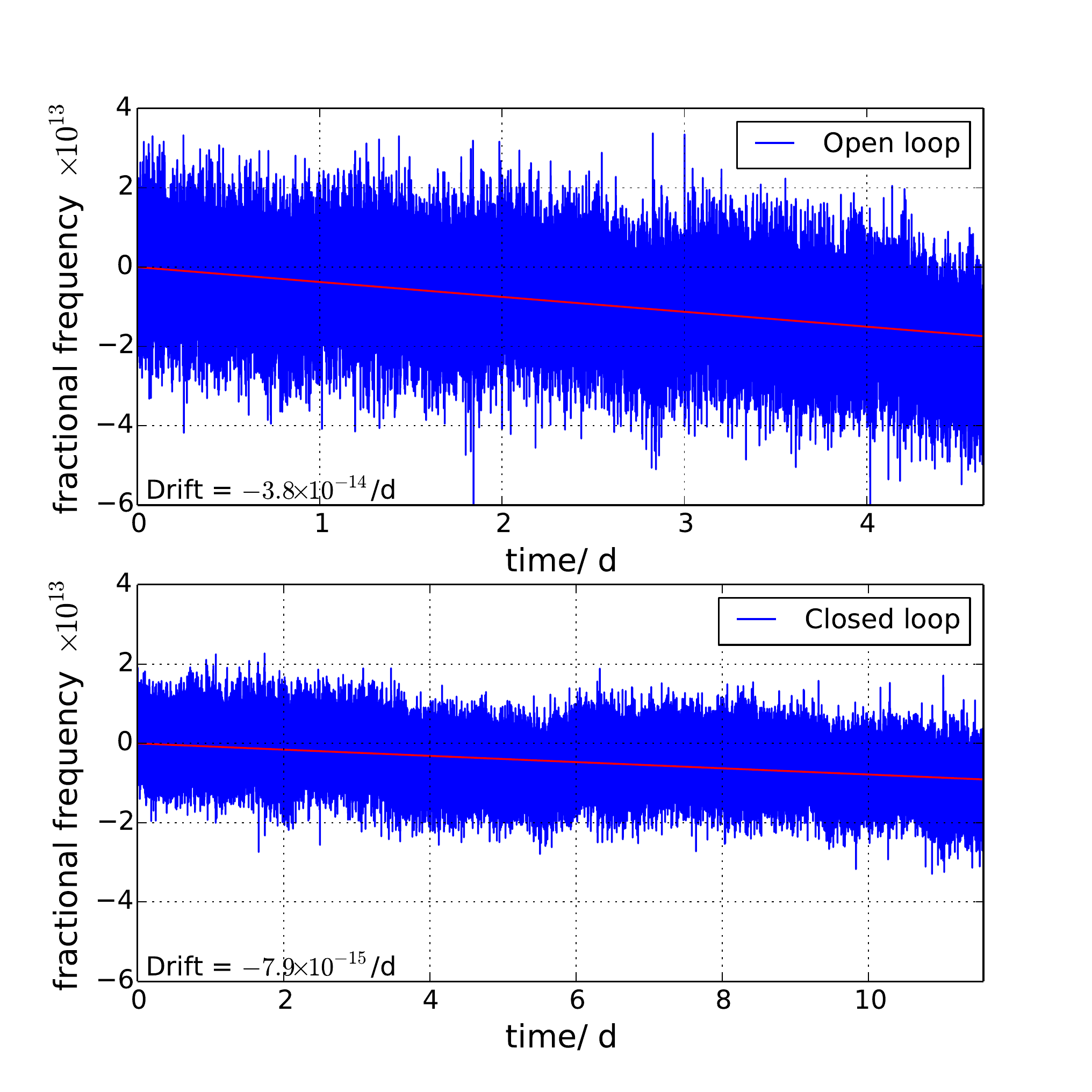}
	\caption{Frequency data of long-run measurements with the microwave field amplitude not stabilized (upper trace, ``open loop'') and with the active stabilization active (lower trace, ``closed loop''). Both data sets are presented with no drift removal. The drift is shown as the (red) trend line.}	
	\label{fig:freqs}
\end{figure}

In \Cref{fig:freqs}, frequency data of two long-run measurements are presented. In the upper trace no active control on the field amplitude is performed (``open-loop'' in the following). In the ``closed-loop'' case, the microwave field amplitude was actively stabilized as described in \Cref{sec:uW_stab}. A linear drift is assumed and calculated as a best fit in the two cases (red thick line). It is around $-4 \times 10^{-14}$/day in the open-loop measurement, while after applying the microwave amplitude stabilization, it remains at the level of $-8 \times 10^{-15}$/day over more than ten days. It is then significantly reduced if compared to the previous measurement.

\begin{figure}[h]
	\centering
	\includegraphics[width=0.45\textwidth]{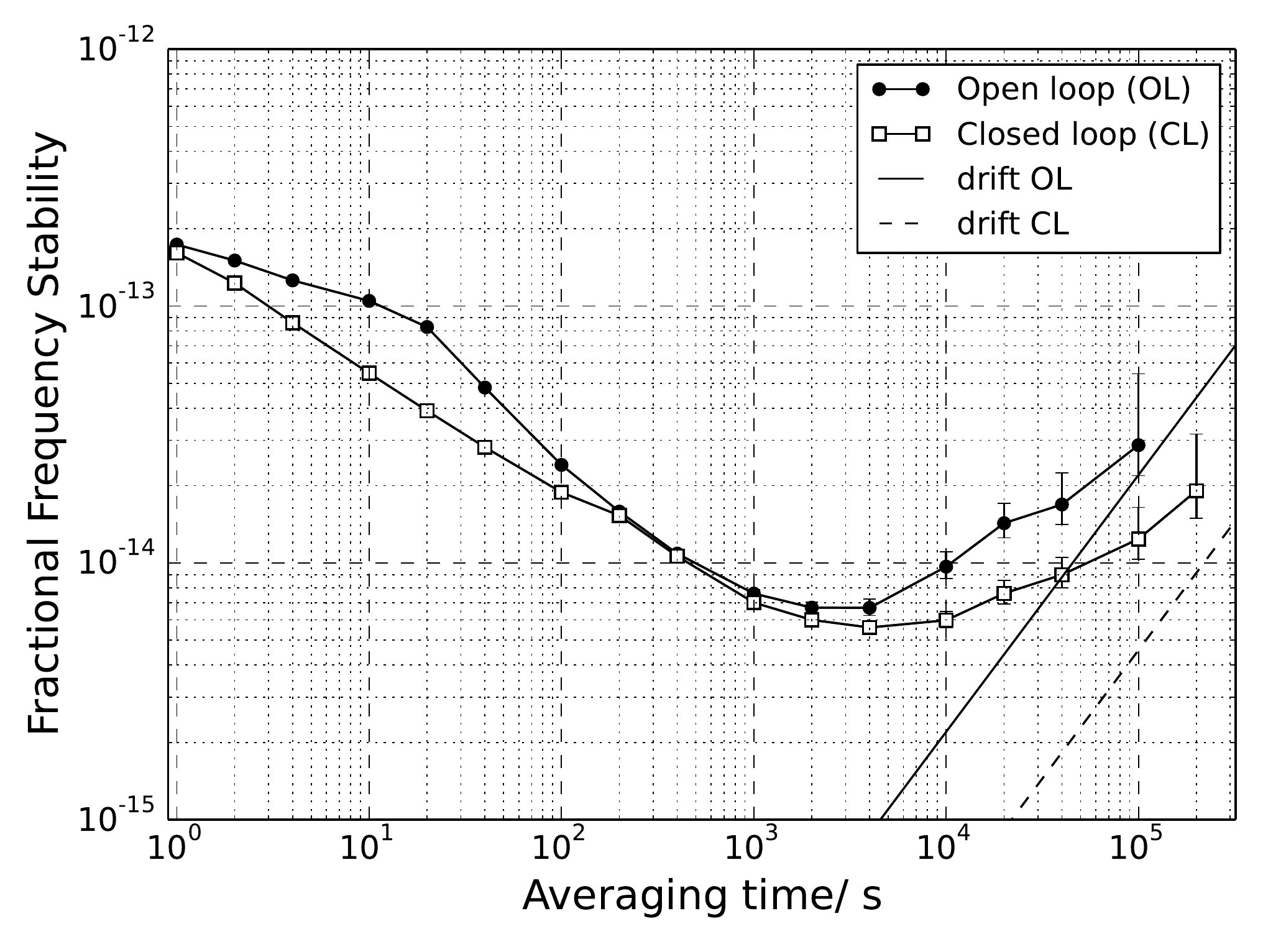}
	\caption{Overlapping Allan deviation for the datasets of \Cref{fig:freqs}. Measurement bandwidth $f_h = $ \SI{0.5}{\hertz}. No drift-removal. The calculated drift contributions are shown as trend lines.}	
	\label{fig:adevs}
\end{figure}

In \Cref{fig:adevs}, the clock overlapping Allan deviations for the same dataset are presented (no drift removed). The bump at 20 s in the first case is due to the reference, while the short-term in the second case is representative of the Rubidium clock only, as the reference instability was brought to a lower level. Regarding the effect of the stabilization on the long-term performances, a significant improvement around averaging times of $10^4$ s and longer can be noticed. Looking at the trend lines representing the drift contributions, we can observe that in closed loop there is an additional instability contribution, apart that coming from the linear drift. We attribute that to other sources of long-term instabilities that show up as the microwave amplitude contribution is reduced. For example in this set-up a temperature instability at cell location of \SI[mode=text]{100}{\micro\kelvin} at \SI[mode=text]{2e5}{\second} can lead to a contribution as large as \num[mode=text]{2e-14}  in fractional frequency stability \cite{Calosso:2012}.

We can then look back at the estimated values for the daily frequency variation due to microwave amplitude fluctuations calculated in the model of \Cref{sec:cav_pull_calc}. When thermal effects directly affecting the physical package are properly treated, we expect the instability contribution from the applied microwave field to be the leading term for a wide range of $Q$ and detuning. By stabilizing the pulse amplitude, the residual sensitivity to cavity pulling comes mainly from the slow change rate of the cavity quality factor. This should keep the cavity pulling contribution to the frequency instability in the $10^{-15}$ region for the usual experimental conditions.

\section{Conclusions}
We presented a technique to actively stabilize the amplitude of the microwave field in compact atomic clocks. The technique we propose in this paper applies to clock working in pulsed regime and exploits the atoms themselves as discriminators, without the need of any external reference. Notably, with the four-step sequence shown in \Cref{tab:sequence} we are able to lock both the frequency and the amplitude of the interrogating microwave to the atoms. All the information is retrieved directly from the clock signal, without using additional hardware.

As extensively discussed in \cite{long_term:2010}, the fluctuations of the microwave power represent one of the main sources of clock instabilities in vapor cell clocks. Specifically, cavity pulling is recognized as the physical phenomenon through which microwave power fluctuations are transferred to the atoms. We demonstrated that our approach can significantly reduce this effect.

Other effects might contribute to limit the clock stability reported in \Cref{fig:adevs}, such as instabilities in the phase of the microwave field and/or thermal instabilities related to the buffer gas \cite{Calosso:2012}. However, the implementation of the microwave amplitude stabilization scheme yields a substantial improvement of the clock stability in our case.

Cavity pulling is assumed to give the main contribution to the medium-long term stability of compact cold atom clocks as well \cite{Liu:2016}. This is mainly due to the fact that in compact devices, differently from atomic fountains, the whole clock operation takes place inside the microwave cavity. The atoms are then sensitive to cavity-related effects during the free evolution phase, where Ramsey fringes are produced. We point out that the stabilization method described in this paper can be extended to cold-atom compact clocks as well.

\label{sec:conclusion}

\appendix
\label{appendix}

We report the complete formula as derived in \cite{cav_pulling:2006} together with the parameters related to our experimental apparatus.
The phase-shift related to cavity-pulling is mainly due to the coherent emission of microwave photons by the atomic coherence, oscillating at the angular frequency $\omega_{12}$ during the free-evolution time. In the presence of a microwave cavity, this coherent emission excites the main resonant cavity mode and acts as a source of magnetic field that will eventually cause a phase-shift in the atomic coherence. This phase-shift is found to be:
\begin{equation}
\phi_0 = \frac{2 Q_L \Delta \omega_C}{\omega_{12}} \ln \lbrace \cosh A(T) + \Delta\cos(\theta)\sinh A(T) \rbrace
\end{equation}
where $\Delta$ is the ground state population inversion after optical pumping, $T$ is the free-evolution time and the term $A(T)$ accounts for the decay of the hyperfine coherence (taking place at a rate $\tilde{\gamma}$):
\begin{equation}
A(T) = \frac{k}{\tilde{\gamma}}\left|\Delta\right| \left(1- \exp(-\tilde{\gamma} T) \right)
\end{equation}
In the previous equation, $k$ is the emission rate of microwave photons per atom multiplied by the filling factor $\eta'$ and is defined as:
\begin{equation}
k = \frac{\mu_0\mu_B^2 Q_L \eta' n}{\hbar (2I+1)}
\end{equation}
where $n$ is the Rb atomic density, $\mu_0$ the vacuum permeability, $\mu_B$ the Bohr magneton, and $I$ the nuclear spin. Finally, the resulting frequency shift observed in the clock frequency is simply:
\begin{equation}
\Delta\omega_{cp} = - \frac{\phi_0}{T}
\end{equation}
For the experimental conditions discussed in \Cref{sec:setup}, we have $\tilde{\gamma} = 210 \;\mathrm{s^{-1}}$, $\eta'=0.31$, $n\simeq 2.4 \times 10^{11}/\mathrm{cm}^3$, $T=3$ ms.

\FloatBarrier

\bibliographystyle{ieeetr}
\bibliography{./biblio_stab_uW}

\end{document}